\documentclass[10pt,english,a4paper]{paper}
\pdfoutput=1

\usepackage[T1]{fontenc}
\usepackage[latin9]{inputenc}
\usepackage{amsthm}
\usepackage{amsmath}
\usepackage{geometry}
\usepackage{lmodern}
\usepackage{babel}
\usepackage{natbib}
\usepackage{nameref}
\usepackage{color}
\usepackage{graphicx}

\title{On the Ambiguity of Interaction and Nonlinear Main Effects in a Regime of Dependent Covariates}
\author{Hannes Matuschek$^{a,b}$ \and  Reinhold Kliegl$^a$}
\institution{$^a$ Department of Psychology, University of Potsdam, Germany \and $^b$ Focus area: Dynamics of complex systems, University of Potsdam, Germany}

\newcommand{\marked}[1]{#1}

\begin{document}

\maketitle


\begin{abstract}
The analysis of large experimental datasets frequently reveals significant interactions that are difficult to interpret within the theoretical framework guiding the research. Some of these interactions actually arise from the presence of unspecified nonlinear main effects and statistically dependent covariates in the statistical model. Importantly, such nonlinear main effects may be compatible (or, at least, not incompatible) with the current theoretical framework. In the present literature this issue has only been studied in terms of correlated (linearly dependent) covariates. Here we generalize to nonlinear main effects (i.e., main effects of arbitrary shape) and dependent covariates. We propose a novel nonparametric method to test for ambiguous interactions where present parametric methods fail. We illustrate the method with a set of simulations and \marked{with reanalyses (a) of effects of parental education on their children's  educational expectations and (b) of effects of word properties} on fixation locations during reading of natural sentences, specifically of effects of length and morphological complexity of the word to be fixated next. The resolution of such ambiguities facilitates theoretical progress.
\end{abstract}

\section{Introduction}
Psychological processes are complex and theoretical advances in an understanding of their dynamics are linked with dissociations of their effects in manifest behavior. Frequently, such dissociations are established with interactions between main effects. Here we show that sometimes such interactions may be due to nonlinear main effects and that it may not be possible to distinguish between alternative explanations. This problem arises \marked{in many areas across the entire spectrum} of psychological research.

In general, we describe a response obtained in \marked{a study or} an experiment as a function of one or more covariates on which the response may depend%
\footnote{In statistical modeling the response is usually called
\emph{dependent variable} and the covariates \emph{independent variables}. In this
article we examine effects in the presence of statistically dependent
\emph{independent variables}. To avoid confusion, we will refer to all 
\emph{independent variables} as covariates and to the \emph{dependent variable}
as the response variable throughout this article.}. The functional dependency of
the response variable on the set of covariates is described in terms of a
statistical model that is fitted to observations. We have a wide variety of statistical models at hand to analyze the observations and to estimate the effects of the covariates on the response. In this article, we use analysis of variance (ANOVA) of factorial covariates, linear models (LM), and linear mixed models (LMM) to demonstrate a little known ambiguity between nonlinear main effects and interaction effects in these statistical models. We also present a simple two-step procedure to detect this possible ambiguity.

The ambiguity leads to effects similar to those observed in \emph{suppressor constellations} 
\citep[e.g., ][]{Lewis1986, Tzelgov1991, Friedman2005}. In a suppressor constellation, a covariate (e.g., $z$) that is actually independent of the response (e.g., $y$) but correlates with a second covariate (e.g., $x$), can improve the fit of a model and may therefore be considered a significant effect. In the following, we demonstrate \marked{with a reanalysis of effects of parental education on their children's educational expectations \citep{Ganzach1997} and a} simple artificial example, that similar effects can be observed between interaction effects and nonlinear main effects if two covariates are correlated. In the \emph{\nameref{sec:math}}  section, we describe the origin of these artifacts in some detail and extend the issue to nonlinear dependencies between covariates. We also show that nonlinear main effects and interaction effects are actually ambiguous if covariates are dependent and that this ambiguity cannot be resolved by a statistical method alone. Rather, the experimental control of these covariates is required to ensure their independence and to resolve the ambiguity. Having established the mathematical context, we propose a simple two-step procedure to test whether there is an ambiguity between nonlinear main effects and interaction effects. In the 
\nameref{sec:results} section, we illustrate the effects of this ambiguity and their detection in a
simulation with artificial examples and in a reanalysis of fixation locations during reading.

\marked{We illustrate the ambiguity between nonlinear main and interaction effects with a brief reanalysis of children's educational expectations as a function of their parents' education \citep{Ganzach1997}. Educational expectations of children (EE) are operationalized with the number of years they expect to complete; mothers' (ME) and fathers' (FE) education is indicated by their highest grade achieved. The data were taken from the 1979 \emph{National Longitudinal Survey of Youth (NLSY)} \citep{nlsy1979}. As \citet{Ganzach1997}, we used only the data of those 7,748 out of 12,686 children, who were living with both parents and whose mother's and father's education were available in the dataset.}

\marked{The two covariates ME and FE are correlated $cor(ME, FE)\approx 0.67$. Correlated covariates such as ME and FE are quite common in psychological research and are not limited to surveys like the NLSY;  they are also found in many experiments where not all covariates are under experimental control.}

\marked{ Following \citet{Ganzach1997}, we analyze the data with two LMs. 
\begin{align}
 EE &= a\,ME + b\,FE + c\,ME\times FE + \epsilon \label{eq:ee1} \\
 EE &= a\,ME + d\,ME^2 + b\,FE + e\,FE^2 + c\,ME\times FE + \epsilon \label{eq:ee2}
\end{align}}

\marked{The first model (Eq. \ref{eq:ee1}) describes educational expectation as a linear function of  mother's and father's education as well as their interaction. In the second LM, we also include quadratic effects in ME and FE, capturing some of the possible nonlinearity of the child's educational expectation as a function of mother's or father's educational status. }
 
\begin{table}[!ht]
\centering
\caption{Parameter estimates, standard deviations and t-values for all effects of the LM (Eq. \ref{eq:ee1}, left panel) and LM (Eq. \ref{eq:ee2}, right panel). } \label{tab:ee}
\begin{tabular}{lcccp{0.7cm}ccc} \hline
  & \multicolumn{3}{c}{Model Eq. \ref{eq:ee1}} & & \multicolumn{3}{c}{Model Eq. \ref{eq:ee2}} \\ \hline
  & Estimate & SE & t-value & & Estimate & SE & t-value \\\hline 
 $ME$ & 0.19 & 0.01& 17.3 & & 0.19 & 0.01 & 17.3 \\
 $ME^2$ & --- & --- & --- & & 0.018 & 0.002 & 7.41 \\
 $FE$ & 0.160 & 0.008 & 19.8 & & 0.168 & 0.008 & 20.6 \\
 $FE^2$ & --- & --- & --- & & 0.014 & 0.002 & 8.90 \\
 $ME \times FE$ & 0.017 & 0.001 & 11.8 & & -0.012 & 0.003 & -3.80 \\ \hline
 \multicolumn{8}{p{14cm}}{\textit{Note: All effects with a t-value larger than 2, $|t|>2$, are considered significant. The $R^2$-values for models \ref{eq:ee1} and \ref{eq:ee2} were $0.197$ and $0.208$, respectively; covariates ME and FE were centered.}} 
\end{tabular}
\end{table}

\marked{Table \ref{tab:ee} summarizes the LM analyses. For the first LM (Eq. \ref{eq:ee1}), one finds the intuitive outcome that the child's educational expectation increases with parents' education (positive linear effects of ME and FE) as well as a positive interaction effect, indicating  overadditive educational expectation if both parents achieved high grades. The second LM (Eq. \ref{eq:ee2}) not only finds significant positive quadratic effects of mother's and father's education on the child's expectation but also a negative interaction effect of parents' education. \citet{Ganzach1997} concluded that the hypothesized negative (i.e., underadditive) interaction effect was masked by an artifact of the inadequate modeling of the main effects (here linear instead of quadratic) and the correlation of the covariates. We will show, however, that even the negative interaction effect found by \citeauthor{Ganzach1997} turns out to be ambiguous as it vanishes if one allows for a flexible description of main effects.}


\marked{To understand the origin of} this ambiguity, one may consider a very simple artificial example: 
Assume we observed $5000$ responses $y_i, i=1,...,5000$ (e.g., response times) in an experiment%
\footnote{A large sample ensures a reliable detection of the artifacts. A much smaller
sample, however, already shows a strong effect on the Type-I error rate and power of 
interaction effects.} and also measured two covariates $x$ and $z$. These covariates reflect the
expected processing difficulty of items presented to the subject. Next, let us assume 
that the covariates $x$ and $z$ are correlated, for example, let $z_i = \frac{x_i}{2} + \frac{\,u_i}{2}$, where
$x_i$ and the unobserved covariate $u_i$ are independent and uniformly distributed in 
the interval $[-1,1]$, that is $x\sim U(-1,1),\,u\sim U(-1,1)$. Consequently $x$ and $z$ 
are correlated with $cor(x,z) \approx 0.70$. Finally, let us assume that the responses were
generated by a very simple but nonlinear process that only depends on $x$ but not on $z$, e.g.
\begin{equation}
 y_i = x_i^2 + \epsilon_i\,, \nonumber
\end{equation}
where $\epsilon\sim\mathcal{N}(0,1)$ is some Gaussian noise. This quadratic relation between the
covariate $x$ and the response $y$ implies that they are actually uncorrelated.

The simplest way to analyze such an experiment might be an ANOVA. In this case,
the continuous covariates are turned into so-called factors. For example, one
may turn the continuous covariate $x$ into a binary variable $X$ with a \emph{negative} level if $x<0$
and a \emph{positive} level if $x\ge 0$. Analogously, the continuous covariate $z$ is turned into the binary
variable $Z$. As the generating process $y = x^2 + \epsilon$ is quadratic in $x$,
one may not expect any effect of the factor $X$ on $y$ as the response $y$ only depends on the
absolute value of $x$. Therefore, the expectation value of $y$ given $x<0$ is the same as the expectation
value of $y$ given $x\ge 0$. As the generating process does not depend on $z$ at all, we may expect to
find neither an effect of $Z$ nor an interaction effect of $X$ and $Z$ on $y$.

\begin{table}
 \centering
 \begin{tabular}{lcclcc} 
 \multicolumn{6}{p{12cm}}{\caption{Summary of ANOVA and LM fit.} \label{tab:exres}} \\ \hline 
 \multicolumn{3}{p{6cm}}{ANOVA} & \multicolumn{3}{p{6cm}}{LM}\\ \hline
 Factor & Mean Sq. & F-value & Term & Estimate & t-value \\ \hline 
 $X$ & $0.44$ & $0.421$  & Main effect: $a$ & $0.0164$ & $0.47$ \\ 
 $Z$ & $0.01$ & $0.006$  & Main effect: $b$ & $-0.0392$ & $-0.79$ \\
 $X\times Z$ & $71.79$ & $69.03$ & Interaction: $c$ & $0.955$ & $15.02$ \\ \hline
 \multicolumn{6}{p{12cm}}{ \emph{\small Note: There is a strong interaction effect between the factors $X$ 
 and $Z$ (ANOVA; left panel) as well as between the continuous covariates $x$ and $z$ (LM; right panel).}} \\ 
 \end{tabular}
\end{table}

\begin{figure}[!ht]
 \centering
 \includegraphics[width=.45\textwidth]{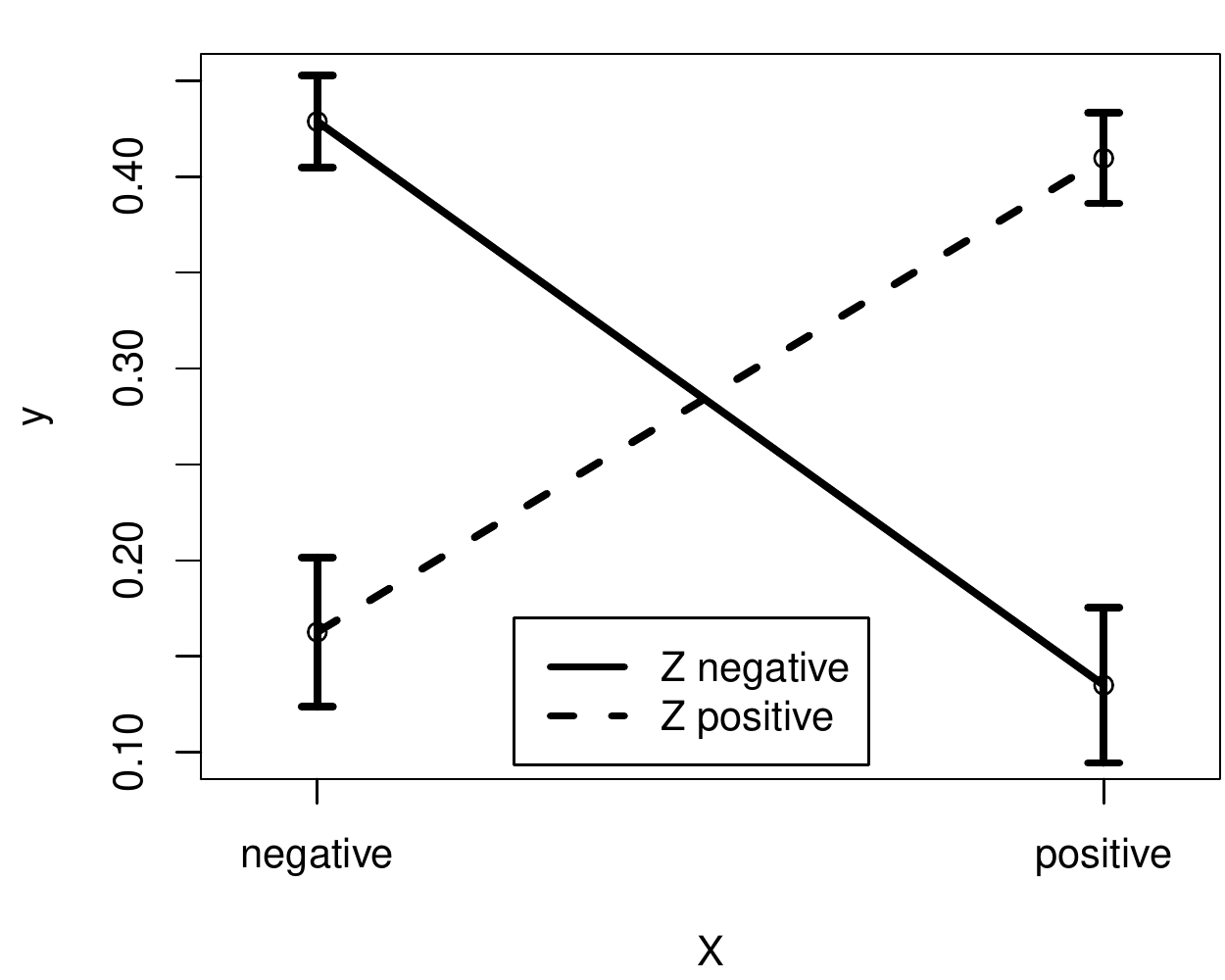} \hfill
 \includegraphics[width=.45\textwidth]{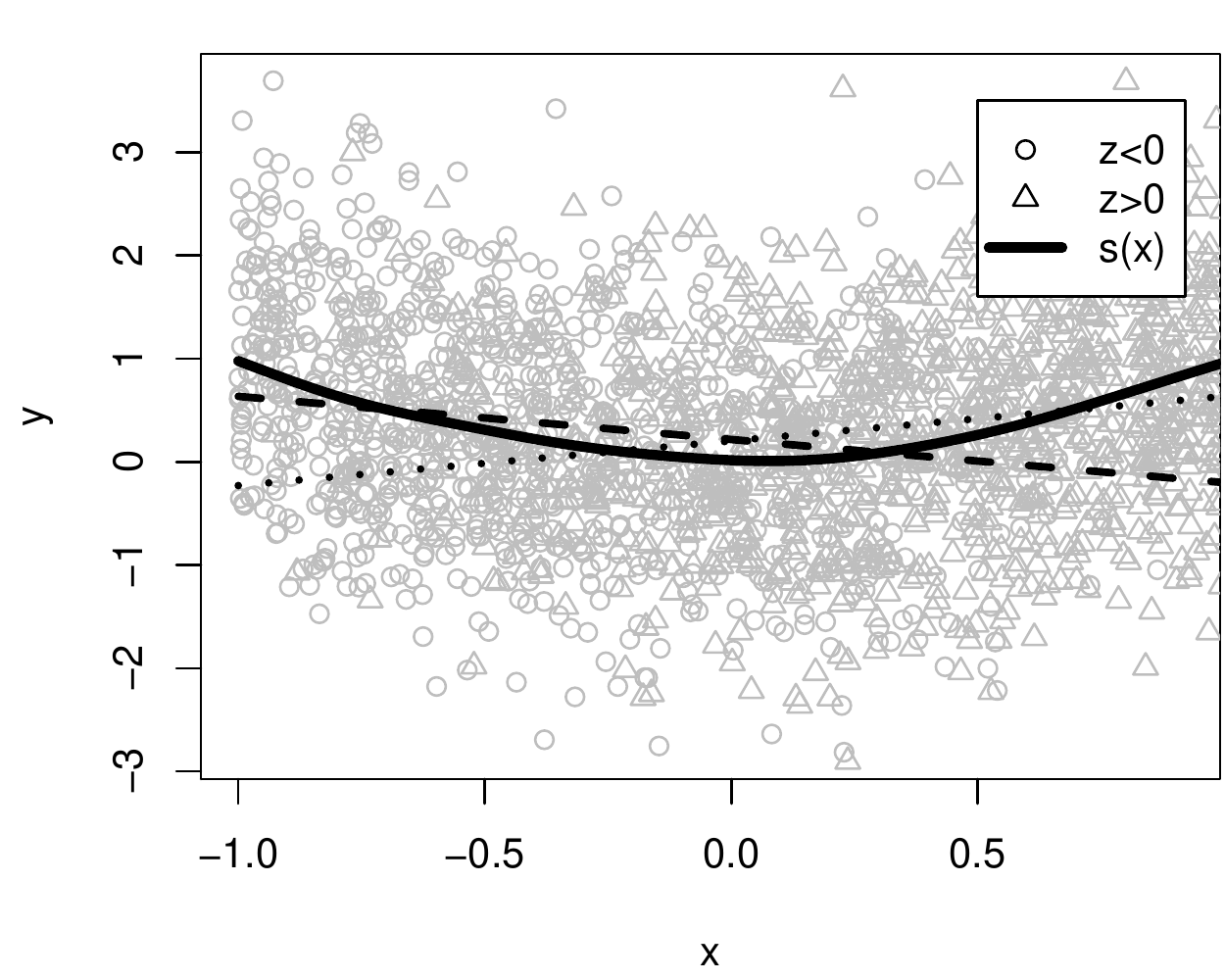}
 \caption{Visualization of the interaction effect. A typical \emph{crossover} interaction is 
 clearly visible for both the ANOVA (left panel) and the LM (right panel).} \label{fig:exvis}
\end{figure}

The left panel of Table \ref{tab:exres} shows a possible outcome of an analysis of the
artificial data. As expected, we do not find strong effects in $X$ and $Z$. Surprisingly, although the generating process does not depend on the covariate $z$ at all, we find a strong interaction effect (see also left panel of Figure \ref{fig:exvis}). This interaction effect can be explained by the nonlinear nature of
the generating process and the correlation between the covariates $x$ and $z$: First, as $x$ and $z$
are correlated, it follows that the probability of $x$ and $z$ having the same sign is increased if $|x|$
is large while it is reduced if $|x|$ is small. This implies that $|x|$ is likely to be small if the levels of
the factors $X$ and $Z$ differ (i.e., $x$ and $z$ have different signs) and that $|x|$ is likely to be large
if the levels match (i.e., $x$ and $z$ have the same sign).  Second, as $y$ depends on $x^2$, it follows
that the expectation value of $y$ will be small if $|x|$ is small and large if $|x|$ is large. Consequently,
the expectation value of $y$ is increased if the levels of the factors $X$ and $Z$ match, while it is
reduced if the levels differ. This is then visible as a typical interaction effect between the factors in the
left panel of Figure \ref{fig:exvis}.

A similar artifact can be found when analyzing the same data with the linear model (LM) 
\begin{equation}
 y_i =  a\,x_i + b\,z_i + c\,x_i\,z_i + \epsilon_i\text{ where } \epsilon\sim\mathcal{N}(0, \sigma^2)\,,\label{eq:lmex}
\end{equation}
where the coefficients $a,b,c$ and the variance parameter $\sigma^2$ are obtained by a
fit of the LM to the data. This LM describes the responses $y$ in terms of a sum of linear functions in the
continuous covariates $x$ and $z$ ($a\,x_i$ and $b\,z_i$, respectively) and an interaction effect 
($c\,x_i\,z_i$). Again, one may expect to find no linear effect of $x$ on the response $y$ as the
generating process is quadratic. One may also expect to find neither a linear effect in $z$ nor an interaction effect as the
generating process does not depend on $z$.

The right panel in Table \ref{tab:exres} summarizes the results of the LM analysis applied
to the same artificial data as used in the ANOVA example. Again, an unexpected strong interaction
effect between $x$ and $z$ can be found. As in the ANOVA example, the interaction effect can be
explained by the correlation of the covariates $x$ and $z$ and the nonlinear dependency of $y$ on $x$:
If $|x|$ is large, the expectation value of $y$ is increased, as well as the likelihood that $x$ and $z$
have the same sign. Consequently, the product $x\cdot z$ is likely to be positive. If $|x|$ is small,
the expectation value of $y$ is reduced, as well as the likelihood that $x$ and $z$ have the same sign.
Consequently, the product $x\cdot y$ is more likely to be negative. Therefore, the product $x\cdot y$
is able to capture at least some part of the quadratic effect of $x$ on $y$ in the generating process 
(see also right panel in Figure \ref{fig:exvis}).
Please note that this artifact is independent of the well known interpretation issues concerning the
scaling and centering of covariates in interaction effects.

This little known issue of interaction effects of dependent covariates was
discussed in \citet{Lubinski1990}, \citet{Cortina1993}, and \citet{Ganzach1997} for linearly
dependent (correlated) covariates and quadratic main effects. In the following section
we generalize this discussion to arbitrarily dependent covariates\footnote{Even if the
covariates were uncorrelated, there may still be nonlinear rather than linear dependencies between
them. Thus, even if covariates are uncorrelated, it is not ensured that these covariates are
statistically independent \citep[see Anscombe's quartet for a beautiful illustration, ][]{Anscombe1973}}
and general
nonlinear main effects. We show that under a regime of dependent covariates,
there exists in fact an ambiguity between interaction and main effects that cannot be
resolved only with recourse to data. Then we introduce a new two-step method to
identify such ambiguous interaction effects, followed by a brief simulation study to 
demonstrate the procedure. Finally, we apply the method to \marked{reanalyses of 
children's educational expectations and} fixation locations during reading.

\section{Mathematical background} \label{sec:math}
In the following we discuss a possible origin of ambiguous interaction
effects that may be found in the analysis of empirical data. For the sake of
simplicity, we will restrict ourselves to simple linear models with polynomial
main effects (i.e., $x$, $x^2$, etc.) and interaction effects between two
mutually dependent covariates (i.e., $x\times z$). The matter, however,
generalizes easily to arbitrary nonlinear main effects and higher-order
interaction effects in more than two variables, as discussed within this section and demonstrated
in the Section \emph{\nameref{sec:sim}} below. Moreover, the restriction on mere 
LMs instead of linear mixed models \citep[LMMs, e.g., ][also introduced below]{Pinheiro2000}
eases the discussion of the mathematical background, as spurious fixed effects
cannot emerge from neglecting significant random effects 
\citep[e.g., ][]{Matuschek2015b, Baayen2015}.

As in the introduction, we assume that $N$ observations of a response variable
$y_i, i=1,...,N$ together with some covariates $x_i$ and $z_i$ are obtained in
an experiment. Further, we assume that $z_i$ depends linearly on $x_i$ as
$z_i = w_x\,x_i + w_u\,u_i$, where $u_i$ is a \emph{hidden} covariate which is
independent of $x_i$ and $w_x\neq 0, w_u\neq 0$. This implies that the
covariate $z$ consists of a part that can be explained in terms of the
covariate $x$ and a part that is independent of $x$. Consequently they
are correlated $\text{cor}(x,z)\neq 0$. 

Now, assuming that the simple nonlinear process
\begin{equation} \label{eq:m1gen}
 y = x^2 + \epsilon\,, 
\end{equation}
generates the responses $y_i$, where $\epsilon \sim \mathcal{N}(0,\sigma^2)$
are independent and identical distributed (i.i.d.). Given the observations 
$y_i, i=1,...,N$ from this generating process,
along with the associated covariates $x_i$ and $z_i$ (please note that the
generating process does not depend on $z$ at all), we then may try to explain
the observations $y_i$ by means of the linear model 
\marked{\begin{equation} \label{eq:m1fit}
 y_i = a\cdot x_i + b\cdot x_i\,z_i + c\cdot z_i + \epsilon_i\,,
\end{equation}}
including an interaction-effect term $x\times z$.

As $z$ depends linearly on $x$, the model (Eq. \ref{eq:m1fit}) can be expanded
to
\begin{align}
  y_i &= a\cdot x_i + b\,x_i\,\underbrace{\left(w_x\,x_i + w_u\,u_i\right)}_{=z_i} + c\underbrace{(w_x\,x_i + w_u\,u_i)}_{=z_i} + \epsilon_i \nonumber \\ 
    &= a\cdot x_i + b\,w_x\cdot x_i^2 + b\,w_u\cdot x_i\,u_i + c\,w_x\cdot x_i + c\,w_u\cdot u_i+ \epsilon_i\,. \nonumber
\end{align}

The expanded model now explicitly contains a quadratic term in $x$,
$b\,w_x\cdot x^2$, which was \emph{hidden} in the interaction effect term
(Eq. \ref{eq:m1fit}). This implies that the interaction effect term
$x\times z$ of the model (Eq. \ref{eq:m1fit}) contains a part that is
implicitly quadratic in $x$. Therefore, given a sufficiently large sample 
size, the fit of the model (Eq. \ref{eq:m1fit}) will report a significant
interaction effect between the covariates $x$ and $z$ (described by $b$
in Eq. \ref{eq:m1fit}), although the generating process (Eq. \ref{eq:m1gen})
neither includes any interactions between these covariates nor depends
explicitly on the covariate $z$ at all. This spurious interaction effect
originates from the improper description of the main effect of $x$ in
the model (Eq. \ref{eq:m1fit}) as a linear one, although the \emph{true} main
effect of $x$ on $y$ is quadratic (Eq. \ref{eq:m1gen}).

Unfortunately, the reverse is also true. The quadratic main effect in $x$ 
may become significant under the following conditions: if the generating process contains an
interaction effect between the dependent covariates $x$ and $z$, such as 
\begin{equation}
y = x\cdot z + \epsilon\,, \label{eq:m2gen}
\end{equation}
and responses from this process are described by quadratic terms in $x$ and
$z$, but without an explicit interaction effect, such as
\begin{equation}
 y_i = a\cdot x^2_i + b\cdot z^2_i + \epsilon_i\,.\label{eq:m2fit}
\end{equation}
This
spurious nonlinear main effect originates from the linear dependence between
the two covariates. As $z = w_x\,x + w_u\,u$, the generating process 
(Eq. \ref{eq:m2gen}) is then equivalent to 
\begin{align}
 y &= x\cdot \underbrace{\left(w_x\,x + w_u\,u\right)}_{=z} + \epsilon \nonumber \\
   &= w_x\,\mathbf{x^2} + w_u\cdot x\,u + \epsilon\,, \nonumber
\end{align}
which implicitly contains a quadratic term in $x$ (in bold above), 
that was once again \emph{hidden} in the
interaction-effect term of the generating process. Although the generating
process (Eq. \ref{eq:m2gen}) is a simple interaction between $x$ and $z$, the
model (Eq. \ref{eq:m2fit}) may report a significant quadratic main effect in
$x$. This implies that in the presence of dependent covariates, an inadequate
model for some given observations may result in an ambiguity between
interaction and nonlinear main effects. 

Of course, if the model matches the generating process, the expected values
of the model parameters will match those of the generating process and hence
allow for a reliable inference about the existence of interaction or
nonlinear main effects. For the analysis of empirical data, however, the
generating process is usually unknown. As cognitive processes are complex,
one may even assume that no model matches the generating process exactly \citep{Box1979}. 
This is usually not a problem, as a chosen model may approximate the generating process 
sufficiently well, at
least in the partial effects of theoretical interest. The presence of mutually
dependent covariates, however, implies an ambiguity between the interaction
effects and nonlinear main effects which can not be resolved. As interaction
effects are typically interpreted in a completely different way than nonlinear
main effects, the ambiguity must be taken into account or at least discussed
in the report.

For simple cases like those above, 
where covariates are only linear dependent (i.e., correlated),
the ambiguity between nonlinear main
and interaction effects can indeed be resolved by including polynomial
main effects up to the same order as the sum of the polynomial degrees of all
interaction effects in the model. For example, a model with one interaction
effect $x\times z$ should also include linear and quadratic main effects in
$x$ and $z$ \citep[e.g. ][]{Lubinski1990, Cortina1993, Ganzach1997}. This
solution, however, is only valid if one can assume that the dependency between
the covariates $x$ and $z$ is linear. For covariates derived from some natural
setting, it is not ensured that the assumption of a linear dependency is justified.  If this
dependency is nonlinear, the ambiguity cannot be resolved and an interaction
effect may then also be explained by a polynomial main effect of a higher order
than $2$ and vice-versa (see Section \emph{\nameref{sec:results}} below).

So far, we have described an effect where spurious interaction effects may appear
in the presence of linear dependent (correlated) covariates and when nonlinear main effects
are inappropriately described in a statistical model (e.g., by linear ones).

If the dependency between the covariates (here $x$ and $z$) is nonlinear 
(e.g., $z = w_x\cdot x^2 + w_u\cdot u$), where $x$ and $u$ are independent, the
description of the data sampled from a generating process containing a cubic
term, for example,
\begin{equation}
 y = x^3 + \epsilon \,, \nonumber
\end{equation}
using a model with an interaction effect between the covariates
$x$ and $z$, for example,
\begin{equation}
 y_i = a\cdot x_i^2 + b\cdot x_i\,z_i + c\cdot z_i^2 + \epsilon_i\,, \nonumber
\end{equation}
leads to a significant interaction effect between $x$ and $z$, as the
interaction effect term $x_i\,z_i=w_x\cdot x_i^3+w_u\,x_i\, u_i$ implicitly contains a
cubic term in $x$.

According to \citet{Lubinski1990}, \citet{Cortina1993}, and \citet{Ganzach1997}, 
the ambiguity between nonlinear main effects and interaction effects should be resolved for the latter model. In fact, this is only the case if the dependency of the covariates $x$ and $z$ is linear. If the dependency is nonlinear, the ambiguity between main and interaction effects reappears.

Alternatively, the dependency between the covariates $x$ and $z$ can be
\emph{removed} by residualizing  a covariate, for example $z$, with respect to
the other covariates (e.g., $x$). The aim of this method is to obtain a new
covariate, $\tilde{z}$, which is independent of all other covariates 
\citep[e.g., ][]{Wurm2014}. Residualization, however, is usually a parametric approach.
That is, residualizing $z$ as $z = a\,x+b\,x^2+\tilde{z}$ will 
ensure that $x$ and $\tilde{z}$ are independent with respect to linear and
quadratic terms. Any higher-order dependency between the covariates $x$ and
$z$ will remain and may still imply an ambiguity of interaction effects between $x$ and the
residualized covariate $\tilde{z}$ and nonlinear main effects. 

Although we restrict ourselves to linear models in the discussion of this
phenomenon%
\footnote{Please note that a similar ambiguity between the main and some interaction
effects may arise in cases where the generating process is an unknown
nonlinear but monotonic mapping of a truly additive process. For example,
$y_i = f(a\,x_i+b\,z_i+c\,x_iz_i)+\epsilon_i$ 
\citep{Garcia-Marques2014}. As the function $f(\cdot)$ is unknown, the
ambiguity between the main and interaction effects cannot be resolved even if
the covariates $x$ and $z$ are independent. This type of interpretation issue
is different from the one discussed here and it can neither be resolved by
ensuring the independence of the covariates nor identified by a non-parametric
description of the main-effect terms.}, the problem also appears in generalized linear (mixed) models 
\citep[e.g., ][]{Demidenko2004a} and even in simple ANOVAs by splitting a continuous variable into two
categories (e.g., $x<0$ and $x\ge 0$), as demonstrated above%
\footnote{Of course, covariates should definitely not be converted to
categorical variables to begin with \citep[e.g., ][]{McClelland2015}.}. The effect will
appear as a significant interaction effect of the associated categorical
variables if the generating model contains a quadratic term 
(e.g., $y = x^2+\epsilon$) and the covariates $x$ and $z$ are linearly dependent.

\section{Detecting ambiguous main and interaction effects}
In the following, we introduce a simple two-step approach to detect a possible ambiguity 
between nonlinear main effects and interaction effects.
In general, neither the exact main-effect functions nor the exact dependencies
between the covariates are known, hence an appropriate description of the main
effects becomes increasingly difficult, especially when relatively
large samples are involved. In these cases, even small discrepancies between the
parametric model and the \emph{true} generating process, as well as weak
dependencies between the covariates, may lead to highly significant interaction
effects. This issue reflects the general limitation of parametric approaches to describe a 
functional dependency of a response variable on covariates.

Therefore, we need a non-parametric and adaptive method for the description of the
main effects that allows for an increasing flexibility in the
description of main effects as more and more data become available. Such an
approach avoids the problem that even a relatively small mismatch
between the parametric model and the \emph{true} main effects yields a significant result,
because the method adapts to the increased sensitivity of the statistics with
increasing sample sizes. 

Splines are a versatile tool. They allow for exactly this adaptive and
nonparametric description of the main-effect functions \citep{Silverman1985}.
They are generically smooth or at least continuous functions in one or more
variables and are obtained by searching for a trade-off between the 
goodness-of-fit to the data and the \emph{wiggliness} (complexity) of the function.
In most cases, the so-called \emph{thin-plate regression spline} can be used. 
A spline is a function $s(x)$, given the data $y_i$ at $x_i, i=1,...,N$, that solves
the optimization problem
\begin{equation}
 s(x) = \underset{f(x)}{\text{argmin}} \left(\sum_{i=1}^N\left| y_i - f(x_i)\right|^2 + \lambda \int_{-\infty}^\infty \left|\frac{\partial^2 f(x)}{\partial x^2}\right|^2\,dx\right) \,, \label{eq:crspline}
\end{equation}
where $\lambda$ is the parameter that determines the trade-off between the
goodness-of-fit ($\sum_i\left| y_i - f(x_i)\right|^2$) and the
\emph{wiggliness} of the function ($\int \left|\partial^2_xf\right|^2\,dx$). 

The close relation between spline estimates and LMs \citep{Kimeldorf1970,
Wahba1990}, not only led to the unification of these two tools into one, the additive
model \citep[AM, e.g., ][]{Wood2006a}, but also allows us to determine the trade-off between
the goodness-of-fit and \emph{wiggliness} ($\lambda$) by means of maximum likelihood 
\citep[e.g., ][]{Wood2006a}. In contrast to an LM, an AM allows for
the description of main effects by means of arbitrary spline functions
instead of parametric terms like polynomials. Recent advances in the
inference methods of AMs \citep{Wood2003, Wood2012, Bates2014} have made it possible
to apply these techniques even to large datasets with many covariates
\citep{lme4, mgcv}. Readers interested in non-parametric spline fits 
may want to consult \marked{ \citet{Wahba1990} for mathematical details or \citet{Wood2006a} and  Appendix \ref{sec:spline} for applied perspectives.}

Of course, the flexibility of AMs does not come without a cost. Although
spline functions in AMs can be considered \emph{nonparametric main effects},
they are actually penalized in order to make the spline regression problem
(Eq. \ref{eq:crspline}, above) uniquely identifiable and to avoid over-fitting
the splines to the data. This penalty towards smoother functions may
introduce a small bias in favor of the completely unpenalized parametric
interaction effects of an AM. This implies that, even if the interaction-effect term
of the AM, $y_i = a\cdot x_i\,z_i + s_x(x_i) + s_z(z_i) + \epsilon_i$ can
be explained entirely by the nonlinear main-effects splines $s_x(x)$ and
$s_z(z)$, the penalty towards smoother functions will introduce a small bias on the
estimate of $a$ and hence increase the interaction effect size. 

In order to reduce this bias, one may prevent the \emph{competition for
variance} between the interaction-effect term and the main-effect splines with
the following two-step procedure.
First, a pure main effects AM \citep[see][for an introduction to AM/AMM fits]{Wood2006a} is fitted to the data:
\begin{equation}
 y_i = s_x(x_i) + s_z(z_i) + \epsilon_i\,. \label{eq:m3main}
\end{equation}
Second, a simple LM containing only the interaction effect term 
of interest is fitted to the residuals of the first model:
\begin{equation}
 \epsilon_i = a\cdot x_i\,z_i + \epsilon'_i\,. \label{eq:m3res}
\end{equation} 

The second model allows us to determine the size of the interaction
effect between the covariates $x$ and $z$, which cannot be explained by
the nonlinear main effects in Eq. (\ref{eq:m3main}); it also allows us
to test whether there remains a significant interaction effect.

Similar to the case
discussed above (Eqs. \ref{eq:m2gen} and \ref{eq:m2fit}), an additive
main-effects model can also explain at least some part of an interaction effect that is
present in the generating model. Therefore, the ambiguity between the
nonlinear main- and interaction-effect sizes remains: 
If nonlinear main effects are not 
adequately modeled, the Type-I error rate of interaction effects might be increased,
while the power to detect existing interaction effects might be reduced as they 
can be described at least partially by nonlinear main effects.

\section{Demonstrations} \label{sec:results}
In this section we demonstrate the effects of the ambiguity described in the 
\emph{\nameref{sec:math}} section and the ability of our two-step procedure to detect
that ambiguity. First, with simple simulations using LMs, we compare the ability to discover ambiguities 
between previous parametric approaches and our non-parametric two-step procedure using AMs.
Second, with \marked{two \emph{real-data} examples}, we demonstrate the ambiguity between an interaction effect and nonlinear main effects with \marked{analyses of children's educational expectations using LMs and} fixation locations during reading of Uighur sentences with linear mixed models (LMMs).

\subsection{Simulations} \label{sec:sim}
In this brief simulation study, we demonstrate some effects of the ambiguity
between nonlinear main effects and interaction effects in the presence of
mutually dependent covariates (here $x$ and $z$) and show that 
our two-step non-parametric approach to detect these ambiguities performs
well where current parametric approaches fail.

For each of the $100$ simulation iterations, the covariates for the evaluation of the 
generating processes are sampled as
following: First, $1000$ independent and identically distributed random values
for $x$ and $u$ are sampled uniformly from the interval $[-1,1]$. Then the
second covariate $z$ is obtained as $z = \frac{x}{3} + \frac{2\,u}{3}$ for
simulations 1--5 and as $z = 4x^2 + u$ for simulations 6 and 7. The first
case induces linear dependent covariates $x$ and $z$, where the
correlation between these covariates is about $\text{cor}(x,z)\approx 0.45$.
The second case leads to uncorrelated but nonlinear dependent covariates. 

Table \ref{tab:sim} shows a summary of the simulation results. In Simulation 1
(first row in Table \ref{tab:sim}), the generating process is 
$y = x^2 + \epsilon$ and the model is 
$y = a\cdot x + b\cdot z + c\cdot x\,z + \epsilon$. The simulation
results show an average t-value $\bar{t}$ for the interaction effect term
$x\times z$ of about $\bar{t}\approx 3.57$, suggesting a highly significant
interaction effect, although the generating process neither includes such an
interaction effect, nor does it depend on the covariate $z$ at all. As described
in the introduction, this spurious interaction effect originates from the linear
dependency of covariates $x$ and $z$ and from the inadequate (incorrect)  model
specification with respect to the \emph{true} nonlinear (here quadratic) main
effect in $x$ that is part of the generating process. In this case, the 
ambiguity between nonlinear main effects and interaction effects results in a
severe increase in the Type-I error rate.

In Simulation 2 (second row in Table \ref{tab:sim}), the model properly
accounts for this nonlinearity by including a quadratic term in $x$. In this
case, the average t-value is about $\bar{t}\approx 0.19$, which agrees with the
absence of a significant interaction effect. 

\begin{table}
 \centering
 \begin{tabular}{cllcc} 
  \multicolumn{5}{p{12cm}}{\caption{Simulation results for different generating processes and models.}  \label{tab:sim}} \\ \hline 
    & Generating Process & Model & $\bar{R^2}$-value & $\bar{t}$-value \\ \hline 
  1 & $y = x^2 + \epsilon$ & $y = a\cdot x +  b\cdot z + c\cdot \mathbf{x\,z} + \epsilon$ & $\approx 0.017$ & $\approx 3.84$ \\ 
  2 & $y = x^2 + \epsilon$ & $y = a\cdot x + b\cdot x^2 + c\cdot \mathbf{x\,z} + \epsilon$ & $\approx 0.084$ & $\approx 0.10$ \\ 
  3 & $y = x^2 + \epsilon$ & $y = s_x(x) + s_z(z) + \epsilon$ & $\approx 0.086$ &  \\
    & & $\epsilon = c\cdot \mathbf{x\,z} + \epsilon'$ & $\approx 0.00092$ & $\approx 0.12$ \\ 
  4 & $y = x\cdot z + \epsilon$ & $y = s_x(x) + s_z(z) + \epsilon$ & $\approx 0.042$ & \\
    & & $\epsilon = c\cdot \mathbf{x\,z} + \epsilon'$ & $\approx 0.012$ & $\approx 3.46$ \\ 
  5 & $y = x^3 + \epsilon$ & $y = a\cdot x + b\cdot x^2 + c\cdot \mathbf{x\,z} + \epsilon$ & $\approx 0.11$ & $\approx -0.069$ \\ 
  6 & $y = x^3 + \epsilon$ & $y = a\cdot x + b\cdot x^2 + c\cdot \mathbf{x\,z} + \epsilon$ & $\approx 0.12$ & $\approx 4.14$ \\
    & $z = 4\,x^2 + u$ & & & \\
  7 & $y = x^3 + \epsilon$ & $y = s_x(x) + s_z(z) + \epsilon$ & $\approx 0.13$ & \\
    & $z = 4\,x^2 + u$ & $\epsilon = c\cdot \mathbf{x\,z} + \epsilon'$ & $\approx 0.00010$ &  $\approx 0.34$ \\ \hline
  \multicolumn{5}{p{12cm}}{\emph{\small Note: The last column shows the approximate mean t-value over $100$
 iterations for the interaction effects of the model (in bold). For each
 iteration of the simulation, each analyzing model was fitted to a set of
 $1000$ independent samples from the generating process. An interaction effect
 is considered significant if the average t-value is greater than two, 
 $\left|\bar{t}\right|>2$.}} \\ 
 \end{tabular} 
\end{table}

In Simulation 3, the observations $y$ are
first described by an AM with two main-effect splines $s_x(x)$ and $s_z(z)$.
A second linear model was then used to check if there remains an interaction
effect between the covariates in the residuals of the AM. The AM approach
allows for arbitrary smooth functions as main
effects, in contrast to the linear models in simulations 1 and 2,
where a parametric approach with polynomial main effects was used. 
The results show that no significant interaction effect is found as the 
AM splines are able to capture the nonlinear main effects.

In Simulation 4, we demonstrate that a \emph{true} interaction effect
between the dependent covariates $x$ and $z$ can still be detected reliably.
The second model, fitted to the residuals of the AM, shows a highly
significant interaction effect that could not be explained by the smooth main
effects in $x$ and $z$. Please note that the average value of the estimated
interaction-effect coefficient $a$ is $\bar{a} \approx 0.45$,
although the interaction-effect coefficient of the generating process is
$a=1$. This indicates that at least some part of the \emph{true}
interaction effect was explained by the main-effect splines. Consequently,
the ambiguity between nonlinear main effects and interaction effects reduces 
the power to detect interaction effects here.

Simulations 5-7 demonstrate the effects of linear and nonlinear
dependencies between the covariates $x$ and $z$. In Simulation 5, the
parametric model correctly states that there is no significant interaction
effect, although the cubic main effect in $x$ of the generating process is not
described properly by the model. As discussed above, this is due to the linear
dependency between the covariates $x$ and $z$. The parametric
model does not contain a term that is able to explain the cubic main effect in
$x$ of the generating process. If, however, the dependency between the
covariates $x$ and $z$ is nonlinear, as is the case in simulations 6 and 7,
the interaction effect in the model becomes significant again (Simulation 6).
In this case, the interaction effect contains a cubic contribution in $x$ and hence
describes the cubic main effect of the generating process as an interaction
effect between $x$ and $z$. This, in turn, increases the Type-I 
error rate for the interaction effect. Using the AM approach (Simulation 7) 
solves this issue, as the main-effect splines will adapt to complex 
shaped functions if sufficient evidence is provided by the data.

\marked{\subsection{Analysis of educational expectations}}
\marked{In this section, we continue with the example from the introduction \citep{Ganzach1997}. We test whether the negative interaction effect (see Eq. \ref{eq:ee2}) is ambiguous using the AM approach introduced here. First we model the dependency of the child's educational expectation as a sum of arbitrary but smooth nonlinear functions of the parents' education (Eq. \ref{eq:ee3}) and then we test whether a significant interaction effect between these covariates is still present in the residuals of the AM (Eq. \ref{eq:ee4}).}

\marked{
\begin{align}
 EE &= s(ME) + s(FE) + \epsilon \label{eq:ee3} \\
 \epsilon &= ME \times FE + \epsilon' \label{eq:ee4}
\end{align}
}

\marked{The results of the LM (Eq. \ref{eq:ee4}) show that there is no significant interaction effect  in the model residuals (Eq. \ref{eq:ee3}) ($M=-0.00198$, $SD=0.00129$, $t=-1.53$, $R^2=0.214$ of AM (Eq. \ref{eq:ee3}) and $R^2<0.001$ of LM (Eq. \ref{eq:ee4})). These results suggest that non-linear main effects can account for the negative interaction effect of the correlated parents' education found in the presence of simple polynomial main effects (here modeled with linear and quadratic  terms). Thus, there is  a strong ambiguity between non-linear main and linear interaction effects.}

\subsection{Analysis of fixation locations during reading}
\marked{In our research on eye-movement control during reading, visual (e.g., word length) and lexical (e.g., word frequency) variables influence where we look and for how long \citep[e.g., ][]{Hohenstein2016}. Usually, these and many other variables are correlated and, as far as reading of natural sentences is concerned, not all of them can be controlled in an experiment. For example, word length and frequency are naturally correlated as shorter words are generally more frequent than longer words. Or more specifically, to foreshadow the example we will use in this section, in some languages, such as in Uighur script, multiple suffixes coding gender, case, or number may be serially tacked onto the root of a noun. Consequently, long words will usually be morphologically more complex (i.e., carry a larger number of suffixes) than short words. \citet{Yan2014} showed that eye movement programs are not only influenced by the length of the next word \citep[i.e., fixations are usually close to the center of words irrespective of their length;][]{Rayner1979}, but also its morphological complexity (i.e., fixations are closer to the beginning of words with multiple suffixes). The theoretical relevance of this result is that programming an eye movement to the next word is not only based on visual information (i.e., the length of the next word, delineated by clearly marked spaces between words), but also by subtle information that requires an analysis of within-word details (i.e., identification of suffixes). Thus, the results suggest that we extract quite a bit of linguistically relevant detail from a word before it is fixated. Obviously, these effects are very small and, moreover, high correlations between variables may dramatically reduce the statistical power to detect them. Matters are even worse if we want to test the interaction between continuous covariates such as word length and morphological complexity \citep{McClelland1993}. Examples such as this motivated our research, but, as mentioned above, the dissociation of subtle effects between correlated variables is a pervasive concern in most areas of psychological research.}

In this section, we test whether such an significant interaction effect can be explained by nonlinear main effects in a regime of dependent covariates. Specifically, we reanalyze fixation locations measured during reading of Uighur script as function of covariates relating to the fixated word \citep{Yan2014}. Forty-eight undergraduate students from Beijing Normal University, all of them native speakers of Uighur, read 120 Uighur sentences. Each subject read half of the sentences in Uighur and the other half in a Chinese translation. The analysis is based only on eye movements during Uighur reading.  The variables of theoretical interest are the length of the words and of their root morphemes, as well as the length and number of suffixes. Word length varied from 2 to 21 letters (M = 7.5, SD = 3.0). The percentages of words with 0, 1, 2, and more than 2 suffixes were 34\%, 38\%, 19\%, and 9\%. The length of the root morphemes varied from 1 to 11 letters (M = 4.7, SD = 1.7) and the total number of letters in suffixes varied from 1 to 15 (M = 3.9, SD = 2.3). The data set comprises a total of 13523 fixations. For further details on the experimental setup and analyses using LMMs we refer to \citet{Yan2014}.

Although linear models are frequently used for the analysis of experimental
data, it is more adequate to resort to a broader class of models, the linear
mixed models \citep[LMMs, e.g., ][]{Pinheiro2000, Bates2014}. By including random effects, 
LMMs can be seen as a generalization of LMs. The random
effects describe the deviance of the response from the ensemble mean (the fixed
effects LM part) for a grouping factor as a sample of a common (normal)
distribution. For example, a random intercept for each individual allows us to
model the deviation of the responses of each subject from the ensemble
mean (described by the fixed effects). Frequently, not only individual
differences are modeled as random effects but also item-specific differences
(e.g., random effects of words and sentences in a reading experiment). 
Analogous to the extension of LMs to LMMs, it is possible to extend additive
models to additive mixed models \citep[AMMs, e.g., ][]{Wood2006a, Matuschek2015}. 
The following reanalysis uses LMMs and AMMs as statistical models, not LMs and AMs.

The Uighur script is well known for its rich usage of up to 5 suffixes. 
As the fixation location $x_l$ within a word depends on the length of the fixated word
$l_w$ as well as on the suffix length $l_s$ of that word \citep{Yan2014}, an
LMM describing the fixation location $x_l$ will contain these two trivially
correlated covariates. The word and suffix lengths are linear dependent, as the
length of a word is simply the sum of the root morpheme and the
suffix length, $l_w = l_r + l_s$.  Indeed, the correlation coefficient
between these two covariates is about $\text{cor}(l_w, l_s)\approx 0.7$ in the
Uighur sentence corpus \citep[USC, ][]{Yan2014}. 

Of course, this direct linear dependency between word and suffix lengths can be resolved easily by replacing the word-length covariate $l_w$ with the length of the root morpheme $l_r$. In terms of cognitive processing, however, word length maps onto visual processing whereas the distinction between the length of the root morpheme and the suffixes requires within-word sublexical processing. Thus, there is a theoretical argument for using word length as a covariate. Moreover, we want to demonstrate spurious interaction effects in a \emph{real-world} example using an obvious linear dependency between two covariates. Replacing the covariate $l_w$ by $l_r$ does not solve the issue of dependent
covariates, because the length of the root morpheme is still correlated with the length of the suffix, $\text{cor}(l_r,l_s)\approx 0.46$. 

Fixation locations in words also depend strongly on the amplitude of the incoming saccade ($a$) \citep{McConkie1988, Engbert2010}, that is the distance of the last fixation location from the beginning of the fixated word. Hence a simple LMM for the analysis of the fixation location $x_l$ could be
\begin{equation}
 x_l \sim \left(a + a^2\right)\times l_w \times l_s + (1|Word) + (1|Sentence) + (1|Subject)\,. \label{eq:ulmm1}
\end{equation}

This LMM (Eq. \ref{eq:ulmm1}) is presented in the model notation of the
\emph{lme4} R package \citep{lme4, R}. It describes the fixation location on
letter $x_l$ and incorporates as fixed effects all possible interactions
between a quadratic polynomial of the incoming saccade amplitude $a$ and the
linear effects of word and suffix lengths of the fixated word. To account for
between-subject and between-item variability, the LMM also includes random
intercepts for the fixated word, sentence, and subject.

Table \ref{tab:ulmm1} (left panel) summarizes the results for the fixed effects obtained for
the LMM (Eq. \ref{eq:ulmm1}). Effects are considered significant (printed in bold) if the
absolute value of their t-value is larger than $2$. In this brief analysis, we find significant main effects of
incoming saccade amplitude ($a$, linear part), word length ($l_w$), and suffix
length of the word ($l_s$). Additionally, we find two significant interaction
effects between incoming saccade amplitude and word length ($a\times l_w$),
as well as between word length and suffix length ($l_w\times l_s$).

\begin{table}[!ht]
 \centering
 \begin{tabular}{lccccccc} 
  \multicolumn{8}{p{13cm}}{\caption{Parameter estimates, standard deviations and t-values for all fixed
 effects of the LMM (Eq. \ref{eq:ulmm1}, left panel) and LM (Eq. \ref{eq:ulm2}, right panel). } \label{tab:ulmm1}} \\ \hline 
  & \multicolumn{3}{c}{Model Eq. \ref{eq:ulmm1}} & & \multicolumn{3}{c}{Model Eq. \ref{eq:ulm2}} \\ \hline
    & Estimate & SE & $t$-value & & Estimate & SE & $t$-value\\ \hline
  $a$             & $-103$    & $3.11$ & $\mathbf{-33.3}$ & & $-0.93$ & $2.79$ & $-0.334$ \\
  $a^2$           & $4.37$    & $3.03$ & $1.44$ & & $-1.24$ & $2.82$ & $-0.440$\\
  $l_w$           & $1.36$    & $0.07$ & $\mathbf{18.9}$ & & $0.001$ & $0.03$ & $0.019$\\
  $l_s$           & $-0.09$ & $0.04$ & $\mathbf{-2.34}$ & & $-0.002$ & $0.01$ & $-0.177$\\
  $a\times l_w$   & $-40.6$   & $4.80$ & $\mathbf{-8.46}$ & & $-35.0$ & $4.42$ & $\mathbf{-7.93}$\\
  $a^2\times l_w$ & $2.83$    & $4.67$ & $0.61$ & & $2.45$ & $4.32$ & $0.567$\\
  $a\times l_s$   & $-0.21$  & $2.46$ & $-0.08$ & & $-0.54$ & $2.28$ & $-0.235$\\
  $a^2\times l_s$ & $1.54$    & $2.44$ & $0.63$ & & $1.33$ & $2.28$ & $0.586$\\
  $l_w\times l_s$ & $0.14$   & $0.05$ & $\mathbf{3.06}$ & & $0.01$ & $0.02$ & $0.379$\\
  $a\times l_w\times l_s$ & $3.64$ & $3.13$ & $1.17$ & & $3.21$ & $2.90$ & $1.10$\\
  $a^2\times l_w\times l_s$ & $-0.658$ & $3.04$ & $-0.22$ & & $-1.25$ & $2.82$ & $-0.443$\\ \hline 
  \multicolumn{8}{p{13cm}}{\emph{Note:  All effects with a t-value larger
 than $2$, $|t|>2$, are considered significant and printed in bold font. The $R^2$-values for models \ref{eq:ulmm1}, \ref{eq:uamm2}, and \ref{eq:ulm2} were $0.486$, $ 0.482$, and $0.00962$, respectively.}} \\
 \end{tabular}
\end{table}

Word length $l_w$ trivially correlates with suffix length
$l_s$. Under the assumption that the main effect of suffix length $l_s$ on
fixation location $x_l$ is nonlinear (please note that the LMM above
describes these main effects as linear), the significant interaction effect
between word length and suffix length ($l_w\times l_s$) might be
ambiguous. As described above, we test these potential nonlinearities in the main effects
by first fitting the AMM (Eq.\ref{eq:uamm2}) to the fixation locations, including generic smooth functions
in $a$, $l_w$ and $l_s$ as main effects. Then we test whether the residuals of this AMM still contain significant
interaction effects which could not be explained by the nonlinear main effects with a simple LM
(Eq. \ref{eq:ulm2}).

\begin{align}
 x_l \sim &\, s_a(a) + s_{l_w}(l_w) + s_{l_s}(l_s) \nonumber \\
   &\, + (1|Word) + (1|Sentence) + (1|Subject) \label{eq:uamm2} \\
 \epsilon \sim &\, \left(a + a^2\right)\times l_w \times l_s \label{eq:ulm2}
\end{align}

The right panel of Table \ref{tab:ulmm1} summarizes the parameter estimates, standard deviations, and
t-values of the fixed effects of the LM (Eq. \ref{eq:ulm2}). Unsurprisingly,
main effects are no longer significant, as they are described by the spline
main-effects in the AMM (Eq. \ref{eq:uamm2}). Concerning the interaction
between incomming saccade amplitude and word length ($a\times l_w$), we
find that only a relatively small part of this effect is
explained by the nonlinear main effects, as it is still highly significant and
its coefficient drops only slightly from about $-40$ to about $-35$ (compare
left and right panels in Table \ref{tab:ulmm1}). The interaction between
word and suffix length ($l_w\times l_s$), however, is
explained completely by the nonlinear main effects of the AMM (Eq.
\ref{eq:uamm2}). According to the LM (Eq. \ref{eq:ulm2}), no significant
interaction effect remains in the residuals of the AMM (Eq. \ref{eq:uamm2}).

\begin{figure} [!ht]
 \centering 
 \includegraphics[width=0.6\textwidth]{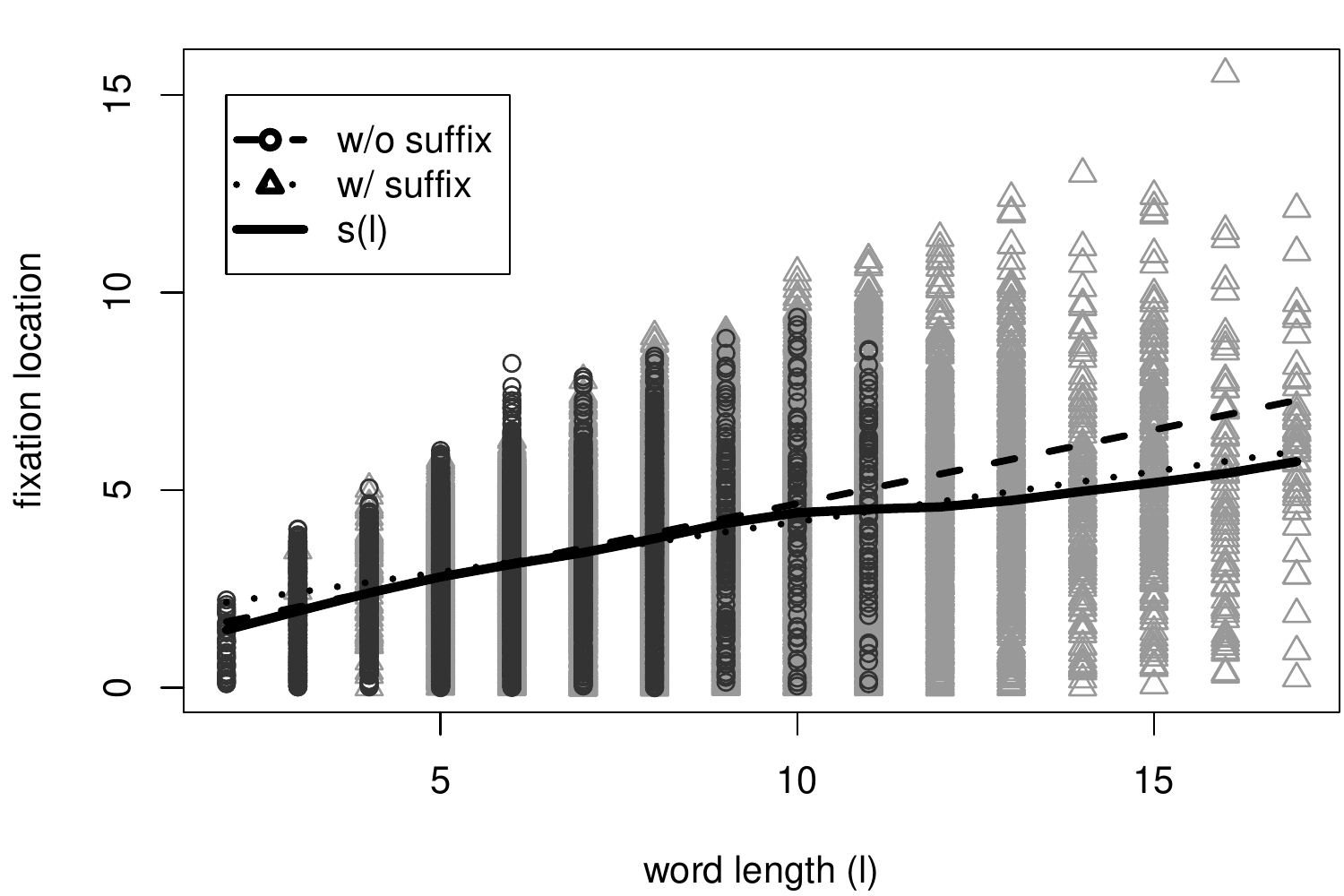}
 \caption{Scatter plots of fixation location and word length for words with and without suffixes (shown as triangles and circles, respectively) and linear main effects of word length on fixation location for words with and without suffixes (dotted and dashed lines, respectively). The solid line shows the spline main effect of word length on fixation location.} \label{fig:uscvis}
\end{figure}

Figure \ref{fig:uscvis} visualizes the ambiguity between a possible nonlinear main effect of word length $l_w$ on fixation location $x_l$ and a possible interaction effect between word length and suffix length $l_s$. This figure shows the scatter plots and linear relationships of fixation locations and word lengths for words without a suffix (circles / dashed line) and words with a suffix (triangles / dotted line). The solid line shows a spline fit describing a nonlinear relationship between word length and fixation location. The scatter plots show that words without suffixes are generally shorter than words with suffixes and the linear trends show that the slopes differ between words with and without suffixes. The latter might be interpreted as an interaction effect between word length and a factor \emph{suffix}. The spline-fit, however, can explain the same effect by allowing for a nonlinear main effect in word length. In fact, the spline appears to follow the linear trend for words without suffixes up to word length $l_w=11$, and to follow the linear trend of words with suffixes for $l_w>11$.

These results, however, do not necessarily imply that the interaction effect does not exist in the first place. The test only informs us that there is a strong ambiguity between the nonlinear main- and interaction-effect terms due to dependencies between these covariates.

\section{Discussion}
We have shown, theoretically, with simulations, and \marked{with reanalyses of children's educational expectations  \citep{Ganzach1997} and of fixation locations in a reading experiment \citep{Yan2014}, that an ambiguity between nonlinear main effects and interaction effects may exist in regimes of dependent covariates. The applications illustrate that this issue shows up in social-science studies and cognitive-science experiments. It may appear in a wide range of analysis tools, such as (G)LMMs, LMs and even simple ANOVAs.} This ambiguity leads to effects similar to what are known as suppressor constellations. However, it extends beyond suppressor constellations in two ways: First, in contrast to suppressor constellations, it is not a variable but rather an interaction effect or a nonlinear main effect that  acts as the \emph{suppressor}. Second, as we extend the dependency between covariates from linear dependent (correlated) to arbitrarily nonlinear dependent covariates, the ambiguity between nonlinear main effects and interaction effects cannot be resolved by statistical analyses alone. To this end, it is impossible to determine which term is the actual \emph{suppressor}. It is therefore crucial to detect whether there is an ambiguity or not.

We propose a novel method to test for this ambiguity, using AMs
that allow for a description of nonlinear main effects by means of generic
smooth functions (splines). These splines are able to \emph{adapt} to the
increase of information often provided by increasing sample sizes; they provide
an appropriate description of arbitrarily complex, but smooth functions. This
adaptive behavior cannot be achieved by resorting to parametric approaches,
such as polynomials of a chosen degree.

Applying this novel method to data from a free-reading experiment of Uighur
script  \citep{Yan2014}, we found that one of two significant interaction effects could be
explained--one could say: completely--as being due to nonlinear main
effects. Regrettably, neither this test nor a non-parametric approach
to the residualization of covariates provides proof of whether or not
an interaction effect is spurious. However, the test is able to detect
ambiguities between possible nonlinear main effects and interaction effects. If
the ambiguity is strong, meaning in cases where an interaction effect can be
explained to a large degree (completely) by nonlinear main effects, then, if at
all possible, an additional experiment should be performed that controls for the
mutually dependent covariates involved in the ambiguous interactions. Only
experimental control of the covariates will resolve the ambiguity.
To this end, we are not yet able to provide a definite threshold for what constitutes a
\emph{strong} ambiguity. This is due to the fact that a change of an interaction effect,
that was significant in an LMM with linear main effects and non-significant in an AMM 
assuming non-linear main effects, is itself not necessarily significant. Moreover, it is
difficult to state the significance of a change in the interaction effect size, as both effect
estimates are based on the same data. Hence, a reliable test-statistic to determine 
whether a significant ambiguity is present will need to be developed.

In our opinion, it is crucial to test whether interaction effects found in an analysis using LMs, LMMs, GLMMs or even ANOVAs involving dependent covariates can be explained by assuming nonlinear main effects. If this is the case, there are potentially strong implications for their interpretation. For example, the Type-I error rate of interaction effects might be increased severely if nonlinear main effects are not modeled adequately. On the other hand, the power of existing interaction effects might be reduced, as they can at least partially be explained by nonlinear main effects. All else being equal, if an interaction effect can be explained completely (compare left and right panels in Table \ref{tab:ulmm1}) by nonlinear main effects, we suggest to consider the interaction effect to be ambiguous, as it can be eliminated from the model by allowing for nonlinear main effects. This represents the usual appeal to parsimony, as nonlinear main effects are usually in-line with theoretical models (as long as they are \marked{monotonic}), while complex interactions usually require much more complex theoretical models.

 \marked{Once we move from quadratic polynomial main effects to splines to account for interactions between main effects, as in the reanalysis of children's educational expectations \citep{Ganzach1997}, one could argue that we are also softening assumptions about monotonic effects; splines are usually wiggly, not strictly monotonic. Including an interaction term might be more parsimonious than the spline-based main effects. However, there are two counterarguments. First, strictly speaking: a quadratic main effect only appears as monotonic as long as it it is restricted to values below or above the minimum or maximum of the function. In this respect, linear and quadratic main effects just as main effects based on splines both represent only an approximate description of a functional relation. Second, if wiggles in main effects replicate, but are small relative to a large monotone declining or increasing trend associated with a main effect, they most likely reflect influences of covariates not yet in the model. Such systematic violations of a monotonic trend may help with the identification of these moderating covariates. Their inclusion will significantly reduce the "model wiggliness".} Our preference, therefore, is that such an ambiguity be resolved with reference to the current state of theory. The specification of a statistical model should be motivated from a theoretical perspective--rather than the hope that the statistical model (i.e., the data analysis) will deliver a theoretical perspective. Thus, we hope to raise awareness of the interpretational options and their  associated constraints. For example, from the current theoretical perspective, the interpretation of the results as an interaction may be preferred. However, future theoretical developments may favor the nonlinear main-effect interpretation. The bottom line is that, ideally, multivariate statistics should be in the service of theory-guided research and not vice versa.

\section*{Acknowledgments}
This research was supported by grant KL 955/6-1 from Deutsche Forschungsgemeinschaft for Research Group 868 \emph{Computational Modeling of Behavioral, Cognitive, and Neural Dynamics}. The simulation and analysis 
\texttt{R} source code as well as the experimental data are available at \emph{Potsdam Mind Research Repository}, \texttt{http://read.psych.uni-potsdam.de/}.

\bibliographystyle{apalike}
\bibliography{references}

\appendix
\marked{\section{Splines and Additive Mixed Models} \label{sec:spline}}
\marked{Splines are versatile tools for non-parametric modeling of arbitrary functions. They approximate some unknown function $g(x)$ of which only a finite number of noisy observations $y_i$ at locations $x_i$ are given, i.e. $y_i = g(x_i) + \epsilon_i$, where $\epsilon_i \sim \mathcal{N}(0,\sigma^2)\,i.i.d.$. Of cause, there are infinite many different functions that would describe the noisy observations perfectly. Hence one assumes a certain penalty $J(f)$ on the function to describe the data. Frequently, the so called \emph{wiggliness} is used as a penalty. That is, the integral over the squared second derivative of the function
\begin{equation}
 J(f) = \int \left|\frac{\partial^2 f}{\partial x^2}\right|^2\,dx\nonumber \,.
\end{equation}}

\marked{This penalty can be considered as a complexity measure. We search for a function $f$ to describe the observed data that minimizes the penalty $J(f)$, which is the least \emph{wiggly} or \emph{bumpy} function. Although the introduction of the penalty $J(f)$ ensures that there is only one function that describes the observations, it may still overfit the data as there exists a single function $f(x)$ that will reproduce the observed data exactly, i.e. $y_i = f(x_i)$. Hence one may search for a trade-off between goodness-of-fit of the function to the data and the penalty $J(f)$ (wiggliness)
\begin{equation}
 \underset{f}{\text{argmin}}\, \underbrace{\Vert y_i - f(x_i)\Vert^2}_{\text{goodness-of-fit}} + \lambda^2 \underbrace{J(f)}_{\text{penalty}}\,. \label{eq:tpsprob}
\end{equation}}

\marked{For this optimization problem, the parameter $\lambda$ determines the trade-off between the goodness-of-fit and the penalty. If $\lambda$ is large, the wiggliness-penalty is strong and a very smooth function will be selected, while for a small $\lambda$ the goodness-of-fit term dominates and a more complex function will be selected. In fact, there is a simple physical analogy to splines and the goodness-of-fit/penalty trade-off. }

\marked{\begin{figure}[!ht]
 \centering
 \includegraphics[width=.45\textwidth]{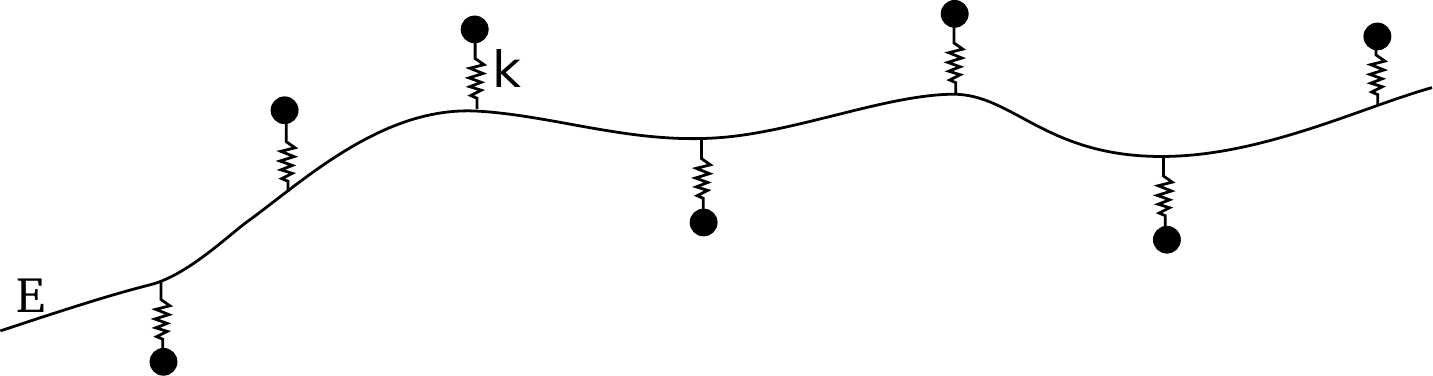}
 \caption{Physical analogy to a spline: An elastic rod suspended in springs. The solution is determined by the stiffness of the springs and rod, as the minimization of the total mechanical energy stored in the bended rod and extended springs.} \label{fig:demo}
\end{figure}}

\marked{Consider a thin elastic rod suspended with springs at some locations in the plane (see Fig. \ref{fig:demo}). These locations represent the observations. The amount by which the springs are extended, represents the goodness of fit and the bending of the rod the \emph{wigglyness} of the solution $f(x)$. The aforementioned trade-off is determined by the ratio of the stiffness of the springs (Hooks constant $k$) and the stiffness of the rod (Young's modulus $E$). Hence, the stiffer the rod compared to the springs, the smoother the solution, while stiffer springs will pull the rod closer to the fixation points (observations) leading to a more wiggly result.}

\marked{Neither the mathematical formulation of the problem (Eq. \ref{eq:tpsprob}) nor its physical analog (Fig. \ref{fig:demo}) allows for a determination of $\lambda$ as the best trade-off between goodness-of-fit and wigglyness. To achieve this, one has to transform the function-approximation problem above (Eq. \ref{eq:tpsprob}) to a statistical regression problem. Hence a theory is required that connects the world of function approximation with the world of statics. The theory of \emph{reproducing kernel Hilbert spaces} (RKHS) does exactly that.}

\marked{A RKHS theorem states that the solution of the optimization problem above (Eq. \ref{eq:tpsprob}) lays within a RKHS uniquely determined by a single function, the reproducing kernel $R(x,x')$, where the optimal solution is given by a finite linear combination of these kernels $f(x) = \sum_i \alpha_i R(x,x_i)$ \citep[e.g., ][]{Matuschek2016, Matuschek2015, Berlinet2004, Wahba1990, Kimeldorf1970}. These kernel functions $R(\cdot,\cdot)$ have another helpful property, they are positive definite functions. This means, they are so-called precision functions\footnote{The precision matrix is the inverse of the covariance matrix and a precision function is a function that generates the precision matrix}. Moreover, the precision matrix $\Gamma_{i,j} = \lambda^2\,R(x_i,x_j)$ is the prior-precision matrix of the spline regression coefficients $\alpha_i$ solving the spline problem above (Eq. \ref{eq:tpsprob}). }

\marked{Within this statistical formulation of the spline regression problem, the unknown parameter $\lambda$ is \emph{just} a prior-precision factor of the regression coefficients, which can be co-estimated by means of maximum likelihood from the data. The ML-estimation of the parameter $\lambda$ also allows the spline to adapt to an increased information about the unknown function being approximated. That is, with more and more data, and therefore more and more information about the unknown function, the $\lambda$ parameter will likely decrease and allow the spline to adapt to smaller features of the unknown function (if there are any). This flexibility of splines is a major advantage over parametric approaches like polynomials and piece-wise defined polynomials (e.g., B-splines), where the complexity of the function fit is determined by the a-priori chosen function class. That is, the degree of the polynomial or the number of knots in a B-spline.}

\marked{Moreover, these regression coefficients $\alpha_i$ of the spline can be considered as random effects of an LMM. This allows to co-estimate regression splines together with \emph{traditional} parametric fixed and random effects of LMMs within a single model, the so-called additive mixed models \citep[AMMs, e.g., ][]{Wood2006a}.}

\marked{Fitting splines to larger datasets, however, gets computationally expensive as the rank of the linear regression problem to be solved is equal to the sample size. Hence Wood developed a provable ideal, low-rank approximation approach to spline regression \citep{Wood2003}. This approach allows to keep the computational cost of a spline fit feasible as sample sizes grow. This approach requires an a-priori choice for the rank of the spline regression problem (the $k$-parameter). Although AMM implementations \citep[e.g., mgcv, ][]{mgcv} provide default values for the choice of $k$, one has to check whether a specific value $k$ is sufficiently large\footnote{Please note that $k$ cannot be chosen too large as long as $k$ is smaller than the sample-size. The computational complexity, however, grows cubic in $k$. To keep the computational costs low, one searches for the smallest $k$ that is still sufficiently large to describe the data.}. In \citep{Wood2006a} a heuristic approach is presented: First a spline with some chosen $k$ (e.g., the default) is fitted to the data. Then a second spline with a doubled rank $2\,k$ is fitted to the residuals of the first spline model. If the latter model finds a significant spline in the residuals of the first, the procedure is restarted with an increased value of $k$ until no significant spline function can be found in the residuals.}

\end{document}